\documentclass[reprint,superscriptaddress,preprintnumbers,nofootinbib,amsmath,amssymb,aps,prd,showkeys,showpacs]{revtex4-2}

\usepackage{graphicx}
\usepackage{dcolumn}
\usepackage{bm}
\usepackage{array} 
\newcolumntype{M}[1]{>{\centering\arraybackslash}m{#1}}
\usepackage{endnotes} 

\usepackage{amssymb}    
\usepackage{color}         
\usepackage{graphicx}     
\usepackage{mathrsfs} 
\usepackage{hyperref} 
\hypersetup{colorlinks=true,linkcolor=blue,citecolor=blue}
\usepackage{ upgreek } 
\usepackage{color,xcolor,fancybox,epsf,rotating,colordvi}
\usepackage{booktabs}
\usepackage{multirow}
\usepackage{fontspec}

\usepackage{ulem}

\newcommand{\GeV}{~\rm GeV}

\newcommand{\fbm}{{~\rm fb}^{-1}}

\newcommand{\abm}{{~\rm ab}^{-1}}

\begin{document}


\title{Testing a 95 GeV Scalar at the CEPC with Machine Learning}

\author{Yabo Dong}
\affiliation{School of Physics and Electronics, Henan University, Kaifeng 475004, China}

\author{Manqi Ruan}
\affiliation{Institute of High Energy Physics, Chinese Academy of Sciences, Beijing 100049, China}

\author{Kun Wang}
\email[Corresponding author:]{kwang@usst.edu.cn} 
\affiliation{College of Science, University of Shanghai for Science and Technology, Shanghai 200093,  China}

\author{Haijun Yang}
\affiliation{State Key Laboratory of Dark Matter Physics, Key Laboratory for Particle Astrophysics and Cosmology (MOE), Shanghai Key Laboratory for Particle Physics and Cosmology (SKLPPC),  School of Physics and Astronomy \mbox{\normalfont\&} Tsung-Dao Lee Institute, Shanghai Jiao Tong University, Shanghai 200240}

\author{Jingya Zhu}
\email[Corresponding author:]{zhujy@henu.edu.cn} 
\affiliation{School of Physics and Electronics, Henan University, Kaifeng 475004, China}


\date{March 2, 2026}

\begin{abstract}
Several possible excesses around $95\,$GeV hint at an additional light scalar beyond the Standard Model.  
We examine the capability of the CEPC to test this hypothesis in the Higgsstrahlung channel  
$e^{+}e^{-}\!\to ZS$ with $Z\!\to\mu^{+}\mu^{-}$ and $S\!\to\tau^{+}\tau^{-}$.  
Full detector simulation shows that the optimal center-of-mass energy to study the 95 GeV light scalar is 210 GeV. 
A deep neural network classifier reduces the luminosity required for discovery by half.
At $L = 20~\mathrm{ab}^{-1}$, the CEPC's $5\sigma$ sensitivity to the signal strength $\mu_{\tau\tau}^{ZS}$ reaches 0.016 and 0.020 for $\sqrt{s} = 210$ GeV and 240 GeV, respectively. The corresponding thresholds for a 5\% precision measurement are $\mu_{\tau\tau}^{ZS} > 0.10$ and $>0.12$. 
At $\sqrt{s}=210$ GeV (240 GeV), $5\sigma$ coverage of all N2HDM-Flipped samples with $\chi^2_{h_{95}}<7.82$ requires $L=800\ \mathrm{fb}^{-1}$ (1.22 $\mathrm{ab}^{-1}$).
These results establish a 210\,GeV run, augmented by machine-learning selection, as the most efficient strategy to confirm or refute the 95\,GeV excess at future lepton colliders.
\end{abstract}

\keywords{new physics, collider phenomenology, CEPC, non-SM Higgs bosons, machine learning}
\pacs{12.60.Fr, 13.66.Fg, 14.80.Cp}

\maketitle
\newpage


\paragraph*{1. Introduction.}
After the discovery of the 125 GeV Standard Model (SM)-like Higgs boson at the Large Hadron Collider (LHC) \cite{CMS:2012qbp, ATLAS:2012yve}, the possibility of additional scalar particles has attracted considerable attention. While the SM contains only one fundamental scalar, there is no symmetry preventing the existence of more. Such extra scalars are theoretically motivated as they may play essential roles in addressing the baryon asymmetry, electroweak phase transition, and dark matter \cite{Sakharov:1967dj, Zhang:1992fs, Kajantie:1995kf}.

Intriguingly, multiple experiments have reported possible local excesses near 95 GeV in different channels, including $b\bar{b}$ at LEP \cite{LEPWorkingGroupforHiggsbosonsearches:2003ing}, $\gamma\gamma$ at CMS and ATLAS \cite{CMS:2024yhz, ATLAS:2024bjr}, and $\tau^+\tau^-$ at CMS \cite{CMS:2022goy}. These excesses can be characterized by the signal strength modifiers \cite{Biekotter:2023oen}: 
\begin{align}\label{Eq:eq1}
    \mu_{bb}^{\exp} &= \frac{\sigma_{\mathrm{BSM}}(e^+e^-\to ZS(\to b\bar{b}))}{\sigma_{\mathrm{SM}}(e^+e^-\to Zh_{95}(\to b\bar{b}))} = 0.117 \pm 0.057\,, \nonumber \\
    \mu_{\tau\tau}^{\exp} &= \frac{\sigma_{\mathrm{BSM}}(gg\to S\to \tau^+\tau^-)}{\sigma_{\mathrm{SM}}(gg\to h_{95}\to\tau^+\tau^-)} = 1.2 \pm 0.5\,, \\
    \mu_{\gamma\gamma}^{\exp} &= \frac{\sigma_{\mathrm{BSM}}(gg\to S\to \gamma\gamma)}{\sigma_{\mathrm{SM}}(gg\to h_{95}\to\gamma\gamma)} = 0.24^{+0.09}_{-0.08}\,,  \nonumber
\end{align}
where $S$ denotes a hypothetical new scalar, and $h_{95}$ is the SM Higgs boson with mass rescaled to 95 GeV. 
It is worth noting that another CMS report on the $S\to\tau\tau$ channel 
in association with top quark pairs~\cite{CMS:2022arx} presents results that differ from the di-tau excess discussed here. As that analysis remains preliminary without subsequent updates, we have focused our discussion on the currently available and more finalized results~\cite{CMS:2022goy}.

These hints have stimulated a broad range of phenomenological studies in the simple extensions of the SM \cite{Kundu:2019nqo, Ge:2024rdr, Aguilar-Saavedra:2020wrj, YaserAyazi:2024hpj, Liu:2018ryo, Wang:2024bkg, Escribano:2023hxj, Borah:2023hqw, Dev:2023kzu, Ahriche:2023hho, Abdelalim:2020xfk, Ashanujjaman:2023etj}, the multi-Higgs model \cite{Biekotter:2019kde,Khanna:2024bah,Benbrik:2025hol, Biekotter:2021qbc, Banik:2023ecr, Aguilar-Saavedra:2023tql, Heinemeyer:2021msz, Biekotter:2021ovi, Biekotter:2022jyr, Biekotter:2022abc, Biekotter:2023jld, Biekotter:2023oen, Arcadi:2023smv, Coutinho:2024zyp, Dong:2024ipo, Banik:2024ugs, Sharma:2024vhv, Dutta:2025nmy, Xu:2025vmy, Benbrik:2024ptw, Benbrik:2022azi, Biekotter:2020cjs, Iguro:2022fel, Azevedo:2023zkg, Belyaev:2023xnv, Coloretti:2023yyq, Maniatis:2023aww}, the supersymmetric models \cite{Cao:2016uwt, Wang:2018vxp, Cao:2019ofo, Hollik:2020plc, Li:2022etb, Domingo:2022pde, Cao:2023gkc, Li:2023kbf, Lian:2024smg, Cao:2024axg, Ellwanger:2023zjc, Ellwanger:2024vvs, Choi:2019yrv, Ellwanger:2024txc}, and the Georgi-Machacek models \cite{Wang:2022okq, Chen:2023bqr, Ahriche:2023wkj, Ghosh:2022bvz, Mondal:2025tzi}.
Ref.~\cite{Drechsel:2018mgd} studies the light Higgs (lighter than 120 GeV) discovery potential at the International Linear Collider (ILC) in the $ S\to b\bar{b}$ decay channel and a center-of-mass energy of $\sqrt{s} = 250\,\text{GeV}$. Ref.~\cite{Wang:2018fcw} gives the exclusion limit on the Z-light Higgs coupling at the ILC in the Higgs-strahlung process with decay channel-independent. Through the study of the $ S\to\tau\tau$ decay channel, Ref.~\cite{Brudnowski:2024iiu} finds that in the search for light Higgs, the search sensitivity for the $S\to\tau\tau$ decay channel is higher than that for the decay channel-independent one.
Ref.~\cite{Sharma:2024vhv} explored the potential of the Circular Electron-Positron Collider (CEPC) to probe this hypothetical scalar through the $ S\to b\bar{b}$ channel, considering a center-of-mass energy of $\sqrt{s} = 250\,\text{GeV}$.

Among the observed anomalies, the possible excess in the $\tau^+\tau^-$ channel is even more pronounced, and in certain beyond the Standard Model (BSM) scenarios, such as the two Higgs doublet model (2HDM)~\cite{Branco:2011iw}, the decay of $S \to \tau^+\tau^-$ can be significantly enhanced. This motivates a dedicated investigation of the $\tau^+\tau^-$ final state. 
The CEPC, primarily designed for precision measurements of the 125 GeV SM-like Higgs boson, is expected to operate at $\sqrt{s} = 240$\,GeV with an integrated luminosity of about 20 ab$^{-1}$~\cite{CEPCStudyGroup:2018ghi, CEPCStudyGroup:2018rmc, An:2018dwb, CEPCStudyGroup:2023quu}. 
However, since a 95 GeV scalar is significantly lighter than the design target of the CEPC, the optimal center-of-mass energy ($\sqrt{s}$) for its discovery remains to be systematically assessed.
Motivated by these considerations, the first part of this work is dedicated to identifying the optimal collision energy for probing a light Higgs with mass near 95 GeV. In addition, we investigate the discovery potential of a light Higgs boson at 95 GeV in the $S\to\tau\tau$ decay channel, employing machine learning techniques such as eXtreme Gradient Boosting (XGBoost), Gradient Boosting Decision Tree (GBDT), and Deep Neural Network (DNN) to enhance the sensitivity. Furthermore, a detailed analysis is performed to evaluate the expected precision of the $Z–S$ coupling and the branching ratio of $S\to\tau\tau$ at the CEPC. Finally, as a concrete example, we apply our approach within the Next-to Two Higgs Doublet Model (N2HDM) to assess its capability in testing specific new physics scenarios.

\paragraph*{2. Monte Carlo simulation.}
We investigate the process $e^+e^- \to Z(\to\mu^+\mu^-)S(\to\tau^+\tau^-)$ at the CEPC. 
The dominant irreducible background is $e^+e^- \to Z(\to\mu^+\mu^-)\tau^+\tau^-$, hereafter denoted $Z\tau\tau$, which contains the same final‐state topology as the signal.  
A sizable reducible background arises from 
$e^+e^- \to Z(\to\mu^+\mu^-)XX$ ($X = j,\,e,\,\mu$), collectively labelled $Zjj$, where one or two $X$ objects can be mis-tagged as hadronic $\tau$ jets. 
Signal and background events are simulated with \textsf{MadGraph5\_aMC@NLO\_v3.4.2}~\cite{Alwall:2011uj, Alwall:2014hca}, interfaced to \textsf{PYTHIA~8.2}~\cite{Sjostrand:2014zea} for parton showering and hadronization, and \textsf{Delphes~3.5.0}~\cite{deFavereau:2013fsa} for detector simulation using the CEPC baseline card with $\tau$-jets identified at an efficiency of 80\%~\cite{Yu:2020bxh}.

Due to the difficulty of reconstructing $\tau$ decays, we employ the recoil mass observable~\cite{Chen:2016zpw}, which allows the invariant mass of the scalar $S$ to be inferred without relying on the reconstruction of its decay products. 
The recoil mass is defined as
\begin{align}
    M_{\mathrm{recoil}} \equiv \sqrt{s + M_{\mu^+\mu^-}^2 - 2\sqrt{s}(E_{\mu^+} + E_{\mu^-})},
\end{align}
and depends only on the well-measured four-momentum of the muon pair. 
This observable is robust against $\tau$-decay ambiguities and provides strong discriminating power between the signal and background. 

\begin{figure}[htb] 
\centering 
\includegraphics[width=1\linewidth]{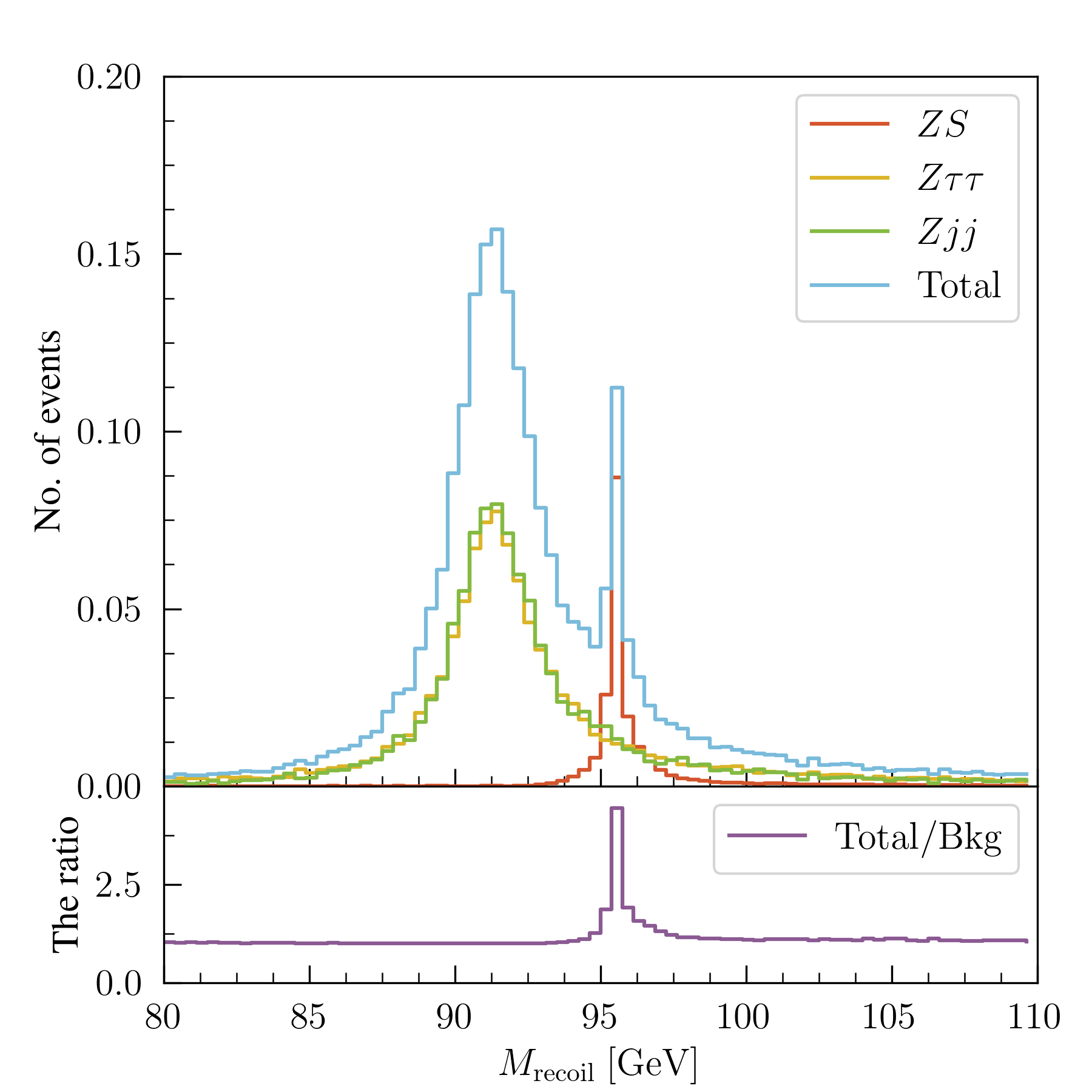}
\caption{\label{f2}
Recoil mass distributions for the signal process $ZS$ with $M_S = 95.5$ (normalized to 0.2), backgrounds $Z\tau\tau$ and $Zjj$ (normalized to 1), and the total (signal + background). 
The lower panel shows the ratio of the total to the background.}
\end{figure}

Fig.~\ref{f2} shows the $M_{\mathrm{recoil}}$ distributions for the signal $ZS$ with $M_S = 95.5$ GeV, the backgrounds $Z\tau\tau$ and $Zjj$, the total (signal + background), and the ratio of the total to the background. 
The number of signal events is normalized to 0.2, reflecting the assumption that the signal cross section is 20\% of the Standard Model prediction, while the background event numbers are normalized to 1.
The signal exhibits a pronounced peak around $M_S$, while both backgrounds show broader distributions peaking near 91 GeV, consistent with the $Z$ mass. 
The total distribution is background-dominated across the mass range, as further reflected in the lower panel, where the ratio of the total to the backgrounds exhibits a localized peak near the signal mass.

We scan the center-of-mass energy $\sqrt{s}$ from 190 to 240\,GeV in 2\,GeV steps and vary the scalar mass $M_S$ from 94 to 100\,GeV in 0.5\,GeV increments. 
For illustration, we take the signal cross section to be $20\%$ of the corresponding SM process with $m_h=M_S$, as computed using \textsf{MadGraph5\_aMC@NLO\_v3.4.2}~\cite{Alwall:2011uj, Alwall:2014hca}.

Event selection requires two identified muons to reconstruct $M_{\mathrm{recoil}}$ and at least one $\tau$-tagged jet to suppress the $Zjj$ background. A second $\tau$-tag is not required, as its efficiency is limited and would significantly reduce signal acceptance despite better background rejection~\cite{Yu:2020bxh}.
Basic kinematic requirements are imposed on the muons:
\begin{align}
    p_{\mathrm{T}}({\mu}) > 10\GeV,\quad p_{T}({\tau}) > 20\GeV, \nonumber \\ 
    |\eta(\mu)| < 2.5,\quad \Delta R(\mu_1,\mu_2) > 0.4, 
\end{align}
where $p_\mathrm{T}$ and $\eta$ denote the transverse momentum and pseudorapidity, respectively, and $\Delta R = \sqrt{(\Delta\eta)^2 + (\Delta\phi)^2}$ measures the angular separation between the muon pair.
Additionally, we require the recoil mass to lie within $|M_{\mathrm{recoil}} - M_S| < 1.5$\,GeV to further suppress the backgrounds.

The basic selection suppresses the reducible $Zjj$ background by roughly four orders of magnitude.  
Requiring $|M_{\mathrm{recoil}}-M_S|<1.5$\,GeV removes about 90\,\% of the remaining $Zjj$ and $Z\tau\tau$ events while retaining $\sim\!80\,\%$ of the signal.

\begin{figure}[htb] 
\centering 
\includegraphics[width=1\linewidth]{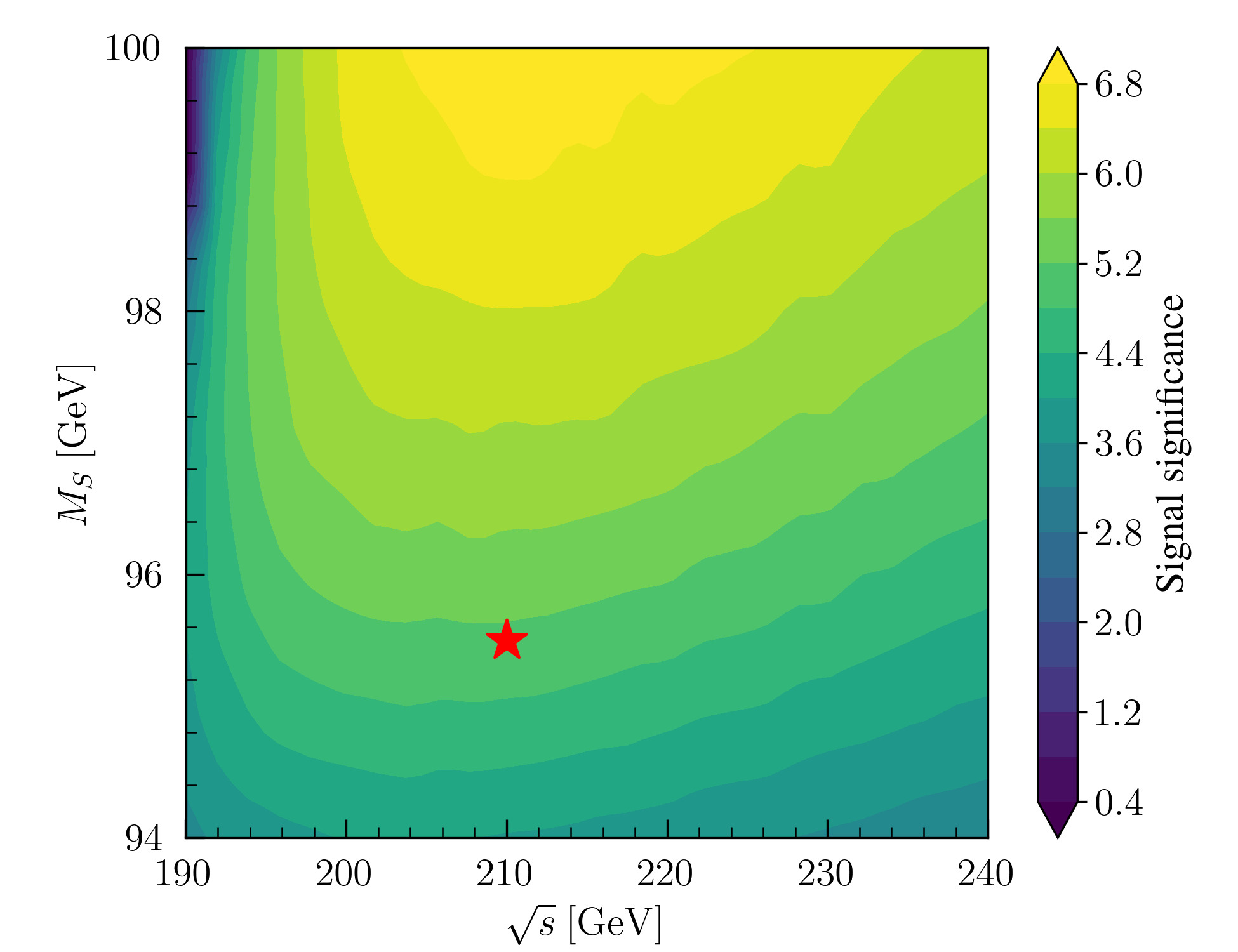}
\caption{\label{f1}
Signal significance $\mathcal{Z}$ in the $M_S$ versus $\sqrt{s}$ plane, assuming an integrated luminosity of 500\,fb$^{-1}$. The red star indicates the benchmark point at $M_S = 95.5$\,GeV and $\sqrt{s} = 210$\,GeV.
}

\end{figure}

Fig.~\ref{f1} shows the signal significance $\mathcal{Z}$ in the $(\sqrt{s}, M_S)$ plane, evaluated for an integrated luminosity of $L = 500$\,fb$^{-1}$. 
The significance is computed using the Poisson expression~\cite{Cowan:2010js},
\begin{align}
    \mathcal{Z}= \sqrt{2L\left[(\mathcal{S}+\mathcal{B})\ln(1+\mathcal{S}/\mathcal{B})-\mathcal{S}\right]},
\end{align}
where $\mathcal{S}$ and $\mathcal{B}$ denote the signal and background cross sections, respectively. 
About one million signal and background events are generated for each parameter point. After applying the cut on $M_{\mathrm{recoil}}$, the dominant SM background originates from $Z\tau\tau$, with approximately 40,000 events surviving. The uncertainty on the expected number of background events is approximately 1\%, which translates to a variation of about 0.4\% in the estimated signal significance. At low statistics, this uncertainty is much smaller than the statistical uncertainty. Furthermore, we have also tested a more conservative scenario by varying the expected number of background events by $\pm 10\%$. The corresponding change in the signal significance is less than 4.5\%.
We observe a moderate increase in $\mathcal{Z}$ with $M_S$, reflecting the more rapid decline of SM backgrounds at higher scalar masses. 
The optimal center-of-mass energy is found to be $\sqrt{s} \simeq 210$\,GeV. At the benchmark point $(M_S = 95.5$\,GeV, $\sqrt{s} = 210$\,GeV), we obtain $\mathcal{Z} \approx 5.1 \sigma$. 
Reaching the same significance at the CEPC design energy of 240\,GeV would require approximately 720\,fb$^{-1}$, about 1.4 times the luminosity needed at 210 GeV.

\paragraph*{3. Machine learning techniques.}
Despite the baseline cuts, the signal remains obscured by irreducible backgrounds, particularly from the $Z\tau\tau$ process.  
To further enhance sensitivity, we apply machine learning (ML) techniques at the benchmark points ($M_S = 95.5$\,GeV, $\sqrt{s} = 210$ and 240\,GeV), comparing three classifiers: gradient-boosted decision trees (GBDT), eXtreme Gradient Boosting (XGBoost), and deep neural networks (DNN).  

A set of 22 kinematic features is used as input, including:
\begin{itemize}
    \item Four-momenta of $\mu_1$, $\mu_2$, and $\tau_1$: $E_{\mu_1,\mu_2,\tau_1}$, $p_x^{\mu_1,\mu_2,\tau_1}$, $p_y^{\mu_1,\mu_2,\tau_1}$, and $p_z^{\mu_1,\mu_2,\tau_1}$.
    \item Transverse momenta and pseudorapidities of $\mu_1$, $\mu_2$, and $\tau_1$: $p_\mathrm{T}^{\mu_1,\mu_2,\tau_1}$ and $\eta_{\mu_1,\mu_2,\tau_1}$.
    \item  Angular separations $\Delta R$ among $\mu_1$, $\mu_2$, and $\tau_1$: $\Delta R[\mu_1, \mu_2]$, $\Delta R[\mu_1, \tau_1]$ and $\Delta R[\mu_2, \tau_1]$
    \item The recoil mass $M_{\mathrm{recoil}}$.
\end{itemize}

The DNN architecture consists of six hidden layers with 64, 48, 32, 24, 16, and 8 neurons, respectively. 
To mitigate overfitting and vanishing gradients, dropout with a rate of 20\% is applied to the first two layers, and each hidden layer uses the Rectified Linear Unit (ReLU) activation function~\cite{Agarap:2018uiz}. 
The output layer contains a single neuron with a sigmoid activation. 
Training is performed using the Adam optimizer with a learning rate of 0.0001. 
Approximately 200k events are used for training and evaluation, with 70\% for training and 30\% for testing. 
The training and testing were performed on an RTX~3060 GPU, and the DNN typically converges within minutes. The model reaches a test accuracy of about 77\% after 10 epochs.
Finally, it reaches a classification accuracy of 78\% (77\%) and an area under the ROC curve (AUC) of 0.87 (0.85) at $\sqrt{s}=210\ (240)$\,GeV. 
Further increasing the size of the training set mainly improves the stability of the training process, while offering limited gains in classification accuracy. Significant additional improvements would likely require more advanced ML architectures or direct access to particle-level information, which we leave for future work.

\begin{figure}[!h] 
\centering 
\includegraphics[width=1\linewidth]{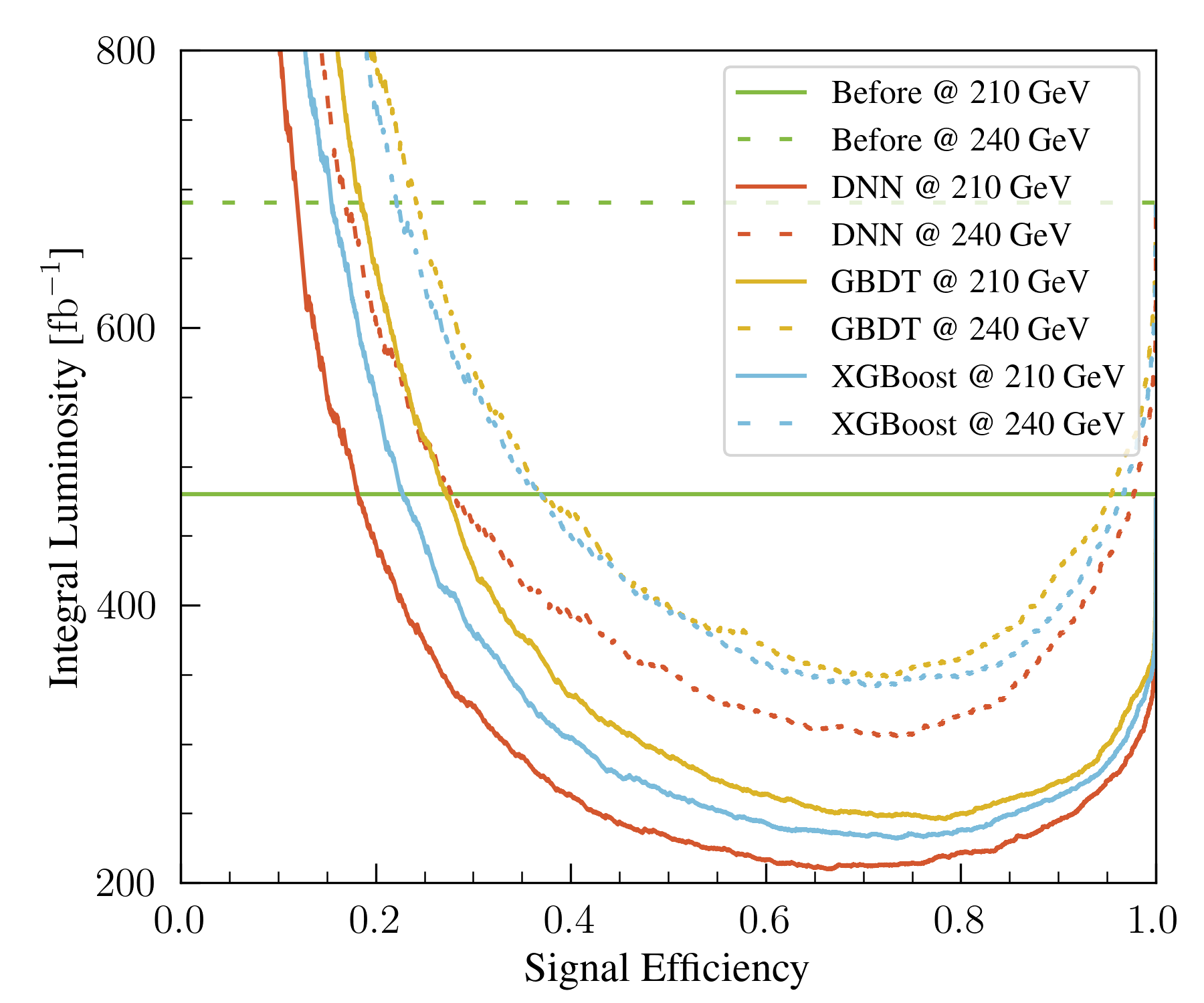}
\caption{\label{f3}
Integrated luminosity required to reach $5\sigma$ significance at the benchmark point, as a function of the signal efficiency. Solid (dotted) curves correspond to $\sqrt{s} = 210\ (240)$\,GeV. 
Colors denote different selection strategies: baseline (before applying ML classifiers, green), DNN (red), GBDT (yellow), and XGBoost (blue).
}

\end{figure}

Fig.~\ref{f3} shows the integrated luminosity required to achieve a $5\sigma$ significance at the benchmark point, as a function of the signal efficiency, before and after applying ML classifiers. 
Here, the signal efficiency is defined as the true positive rate (TPR),
\begin{equation}
   \mathrm{TPR} = \frac{N_{\mathrm{TP}}}{N_{\mathrm{TP}}+N_{\mathrm{FN}}} ,
\end{equation}
where $N_{\mathrm{TP}}$ is the number of signal events correctly classified 
and $N_{\mathrm{FN}}$ is the number of signal events misclassified as background. 
This quantity characterizes the efficiency of the ML classifier in recognizing 
signal events.
The green, red, yellow, and blue curves (dotted for $\sqrt{s} = 240$\,GeV) correspond to the baseline selection (before applying ML classifiers), DNN, GBDT, and XGBoost, respectively, at $\sqrt{s} = 210$\,GeV. 
At each energy, the DNN consistently requires the lowest luminosity for any given signal efficiency. 
The minimum required luminosity is attained at a signal efficiency of approximately 0.66 (0.74) for $\sqrt{s} = 210\ (240)$\,GeV with DNN, reducing the required luminosity from $480\ (690)\;\mathrm{fb}^{-1}$ in the baseline to less than $210\ (310)\;\mathrm{fb}^{-1}$.  
The full cut flow with $L = 500\;\mathrm{fb}^{-1}$ is summarized in Table~\ref{t01}.

\begin{table}[htbp]
  \centering

  \caption{Cut flow of signal and background events (in units of events per 500~fb$^{-1}$) for $\sqrt{s} = 210$ (240)~GeV. Numbers in parentheses refer to the 240~GeV case. The final column shows the resulting signal significance $\mathcal{Z}$.}
\begin{tabular}{@{\hspace{5pt}}ccccc@{\hspace{5pt}}}
    \toprule
    Cuts    & $ZS$ & $Z\tau\tau$ &$Zjj$ &$\mathcal{Z} [\sigma]$\\
    \midrule
    Initial  & 134 (109)  & 1819 (1769) & 24187 (21745) & 0.83 (0.71)\\
    Basic& 59.7 (52.7) & 780.2 (816.7)& 49.4 (4.7)& 2.0 (1.8) \\
    $M_{\mathrm{recoil}}$ & 47.4 (38.7) & 66.0 (66.0)& 5.2 (4.7)& 5.1 (4.3)\\
    ML & 31.5 (28.6) & 8.2 (11.5)& 0.6 (0.8)& 7.7 (6.4) \\
    \bottomrule
\end{tabular}
\label{t01}
\end{table}

\begin{figure*}[!tbh] 
\centering 
\includegraphics[width=0.8\linewidth]{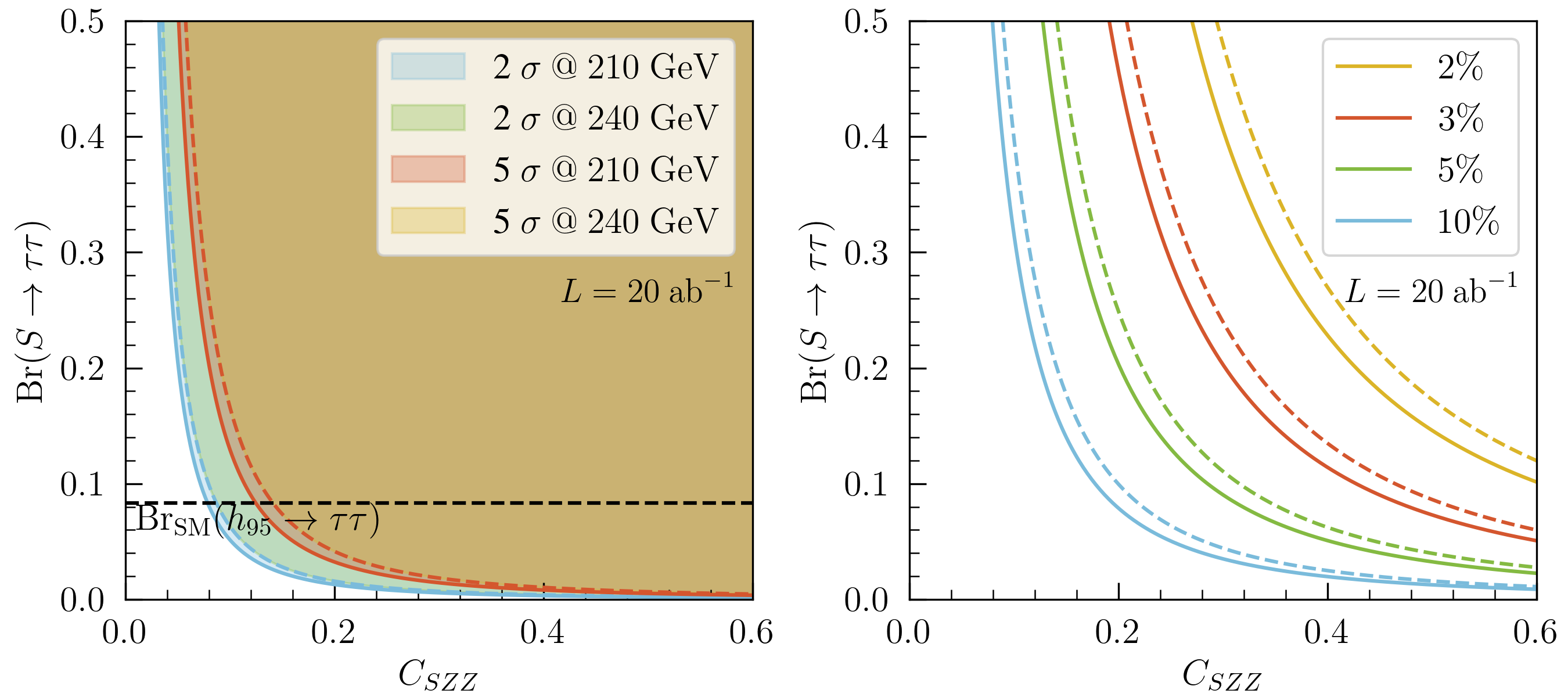}
\caption{\label{f4}
CEPC sensitivity in the 
$\mathrm{Br}(S\to\tau^+\tau^-)$ versus $C_{SZZ}$ planes. 
The left panel shows the expected $2\sigma$ and $5\sigma$ coverage, and the right panel shows the relative statistical precision on the signal yield. Results are presented for $\sqrt{s}=210$ GeV (solid) and $\sqrt{s}=240$ GeV (dashed), assuming an integrated luminosity of $L=20~\mathrm{ab}^{-1}$.}

\end{figure*}

Fig.~\ref{f4} shows the CEPC coverage (left panel) and detection precision (right panel) in the $\mathrm{Br}(S\to\tau^+\tau^-)$ versus $C_{SZZ}$ planes, for $\sqrt{s} = 210$ GeV (solid line) and $240$ GeV (dashed line), with an integrated luminosity of $L = 20~\mathrm{ab}^{-1}$.
Here $C_{SZZ}$ denotes the reduced coupling of $SZZ$ in the N2HDM-Flipped to its SM value.
The left panel presents the expected $2\sigma$ and $5\sigma$ significance contours.  
While the right panel displays the relative statistical precision on the signal yield, evaluated as
\begin{equation}
\delta_{\rm stat}=\frac{\sqrt{(\mathcal S+\mathcal B)L}}{\mathcal S L},
\end{equation}
where $\mathcal S$ and $\mathcal B$ denote the signal and background cross sections. 
The primary sources of systematic uncertainty include $\tau$-jet reconstruction, jet energy scale and resolution, and theoretical uncertainties in the cross-section and shape predictions.
Based on the CEPC Technical Design Report and related studies~\cite{CEPCStudyGroup:2025kmw, Ma:2024qoa, Sharma:2024vhv}, we estimate the overall systematic uncertainty to be about 2\%.
Therefore, the statistical uncertainty dominates until it drops below this level.
Since the production rate in the Higgsstrahlung channel 
$e^+e^-\to ZS(\to\tau^+\tau^-)$ is controlled by both the 
$ZS$ coupling and the branching fraction of $S\to\tau^+\tau^-$, it is convenient to introduce the signal strength
\begin{align}
    \mu_{\tau\tau}^{ZS} \equiv C_{SZZ}^2 \cdot \frac{\mathrm{Br}(S\to\tau^+\tau^-)}{\mathrm{Br}_{\text{SM}}(h_{95} \to \tau^+\tau^-)}.
\end{align}
This single parameter encapsulates the combined dependence on 
$C_{SZZ}$ and $\mathrm{Br}(S\to\tau^+\tau^-)$, and will be used to present the CEPC sensitivity.
In this parameterization, CEPC can probe $\mu_{\tau\tau}^{ZS} > 0.016$ ($0.020$) at $5\sigma$, and $> 0.006$ ($0.008$) at $2\sigma$ for $\sqrt{s} = 210$ ($240$)\,GeV. 
The right panel shows that percent-level precision on $\mu_{\tau\tau}^{ZS}$ can be achieved 
once this quantity reaches $\mathcal{O}(10^{-2})$. 
The corresponding values of $\mu_{\tau\tau}^{ZS}$ required to obtain selected precision targets are summarized in Table~\ref{tab:mu_targets}.
Furthermore, Ref.~\cite{Dong:2024ipo} indicates that the HL-LHC, with an integrated luminosity of 3~ab$^{-1}$, could reach a $5\sigma$ discovery for the process $pp\to t\bar{t}S(\to \gamma\gamma)$ provided the cross section exceeds around 0.3~fb.
In addition, a 5~ab$^{-1}$ CEPC run is projected to be able to cover the current $\sim 2\sigma$ excess at the $5\sigma$ level through the channel $e^+e^-\to Z(\mu^+\mu^-)S(\to b\bar{b})$~\cite{Sharma:2024vhv}.
Together with our present study, these projections outline a promising and complementary path toward probing the potential BSM origin of the 95~GeV resonance at future colliders.
\begin{table}[h]
\centering
\caption{Signal strength $\mu_{\tau\tau}^{ZS}$ required to reach the
indicated relative statistical precision on the signal yield at
$\sqrt{s}=210$ and $240$ GeV with $L=20~\mathrm{ab}^{-1}$.}
\label{tab:mu_targets}
\begin{tabular}{c@{\hspace{45pt}}c@{\hspace{45pt}}c}
\toprule
Target & Required $\mu_{\tau\tau}^{ZS}$& Required $\mu_{\tau\tau}^{ZS}$\\
precision & (210 GeV) & (240 GeV) \\
\midrule
10\% & 0.04 & 0.05 \\
5\%  & 0.10 & 0.12  \\
3\%  & 0.22  & 0.26  \\
2\%  & 0.44  & 0.52  \\
\bottomrule
\end{tabular}
\end{table}

\paragraph*{4. N2HDM-Flipped.}
To demonstrate the broader applicability of our ML approach, we examine the CEPC’s sensitivity to the parameter space of the flipped Next-to-2HDM (N2HDM-Flipped). The same strategy can also be straightforwardly applied to other models with an extended Higgs sector, such as the next-to-minimal supersymmetric model (NMSSM).
N2HDM-Flipped extends the SM with a second Higgs doublet and a real singlet, where the Yukawa coupling of charged leptons is "flipped" from down- (in Type-II N2HDM or NMSSM) to up-quark-like \cite{Chen:2013jvg, Drozd:2014yla, Muhlleitner:2016mzt}. 
A complete definition of the N2HDM-Flipped is given in Section II of Ref.~\cite{Biekotter:2019kde}, detailing its particle content, coupling structure, and differences from scenarios of other Types.
Ref.~\cite{Biekotter:2022jyr} shows that the N2HDM-Flipped scenario can accommodate the excesses observed in several channels. 
The present work further evaluates the CEPC detection potential for the surviving parameter space of this model.

After electroweak symmetry breaking, there are three neutral CP-even states $H_{a,b,c}$, a pseudoscalar $A$, and a charged pair $H^\pm$ in the N2HDM-Flipped.
We identify $H_a$ with the 125 GeV SM–like Higgs boson and take $H_b$ as the putative 95 GeV resonance.
11 new parameters that need to be defined, including the mass of $H_{a,b,c}$, $A$, and $H^\pm$; the ratio of the vacuum expected values (VEV) of two Higgs doublets tan$\beta$; the effective couplings of $H_a$ to massive gauge bosons ($C_{H_aVV}^2$) and top-quarks($C_{H_att}^2$); the mixing matrix elements between $H_{a,b}$ and the singlet field ($R_{a3}^{{\mathrm{in}}}$ and $R_{b3}^{\mathrm{in}}$); the soft $Z_2$-breaking parameter $m_{12}^2$; the singlet VEV $v_S$. 
The 11 independent parameters are scanned in the ranges:
\begin{align}
  95 &< m_{H_b} < 96~\text{GeV},     && \text{sign}(R^{\text{in}}_{a3}) = \pm1, \nonumber\\
  300 &< m_{H_{c},A} < 1500~\text{GeV}, & 580 &< m_{H^\pm} < 1500~\text{GeV}, \nonumber\\
  0.8 &< \tan\beta < 10,            & -1 &< R^{\text{in}}_{b3} < 1, \label{eq:scan}\\[2pt]
  0.70 &< C_{H_aVV}^2 < 1.00,       & 0.70 &< C_{H_att}^2 < 1.20, \nonumber\\
  10^{-3} &< m_{12}^2 < 5\times10^{5}~\text{GeV}^2, & 1 &< v_S < 3000~\text{GeV}. \nonumber
\end{align}

For each parameter point, we impose theoretical consistency conditions, including perturbative unitarity, vacuum stability, and perturbativity, as well as current experimental constraints from Higgs, flavor, and electroweak precision observables. The parameter scan is performed using \textsf{ScannerS\_v2.0.0}~\cite{Coimbra:2013qq, Muhlleitner:2020wwk}, which interfaces with \textsf{HiggsBounds\_v5.10.0}~\cite{Bechtle:2008jh, Bechtle:2020pkv}, \textsf{HiggsSignals\_v2.6.0}~\cite{Bechtle:2013xfa, Bechtle:2020uwn}, and \textsf{SuperIso\_v4.1}~\cite{Mahmoudi:2007vz, Mahmoudi:2008tp} to evaluate physical observables and apply the relevant constraints~\cite{Peskin:1990zt, Peskin:1991sw, Kanemura:1993hm, Akeroyd:2000wc, Arhrib:2000is, ElKaffas:2006gdt, Muhlleitner:2016mzt, ParticleDataGroup:2022pth}.
This work uses a profiled likelihood ratio test with the SM as the alternative hypothesis, defining $\Delta\chi^2 = \chi^2_{\mathrm{N2HDM}}-\chi^2_{\mathrm{SM}}$ and requiring $\Delta\chi^2 < 6.18$ at 95\% confidence level, where $\chi^2_{\mathrm{N2HDM}}$ and $\chi^2_{\mathrm{SM}}$ are given by HiggsSignals to evaluate the compatibility of the 125 GeV Higgs with the experimental data \cite{Bechtle:2020uwn}. 

We further perform a $\chi^2_{h_{95}}$ analysis to evaluate the compatibility of the N2HDM-Flipped with the observed excesses in the $\tau^+\tau^-$, $b\bar{b}$, and $\gamma\gamma$ channels.
This is not a discovery test, but rather a goodness-of-fit check to see whether the model is compatible with the existing experimental hints.
The $\chi^2_{h_{95}}$ is defined as
\begin{align}
    \chi^2_{h_{95}} = \sum_{xx=\tau\tau,\,\gamma\gamma,\,bb} \frac{(\mu_{xx} - \mu_{xx}^{\mathrm{exp}})^2}{(\Delta \mu_{xx}^{\mathrm{exp}})^2},
\end{align}
where $\mu_{xx}$ denotes the model prediction and $\mu_{xx}^{\mathrm{exp}} \pm \Delta \mu_{xx}^{\mathrm{exp}}$ are the experimentally measured signal strengths and uncertainties, respectively, as specified in Eq.~(\ref{Eq:eq1}). 
The uncertainties of the three excesses are treated at the $1\sigma$ level, and for three degrees of freedom the requirement $\chi^2_{h_{95}} < 7.82$ corresponds to consistency with the experimental results at the 95\% confidence level, indicating that the model is not excluded by current data within this statistical tolerance.

\begin{figure*}[tbh] 
\centering 
\includegraphics[width=0.8\linewidth]{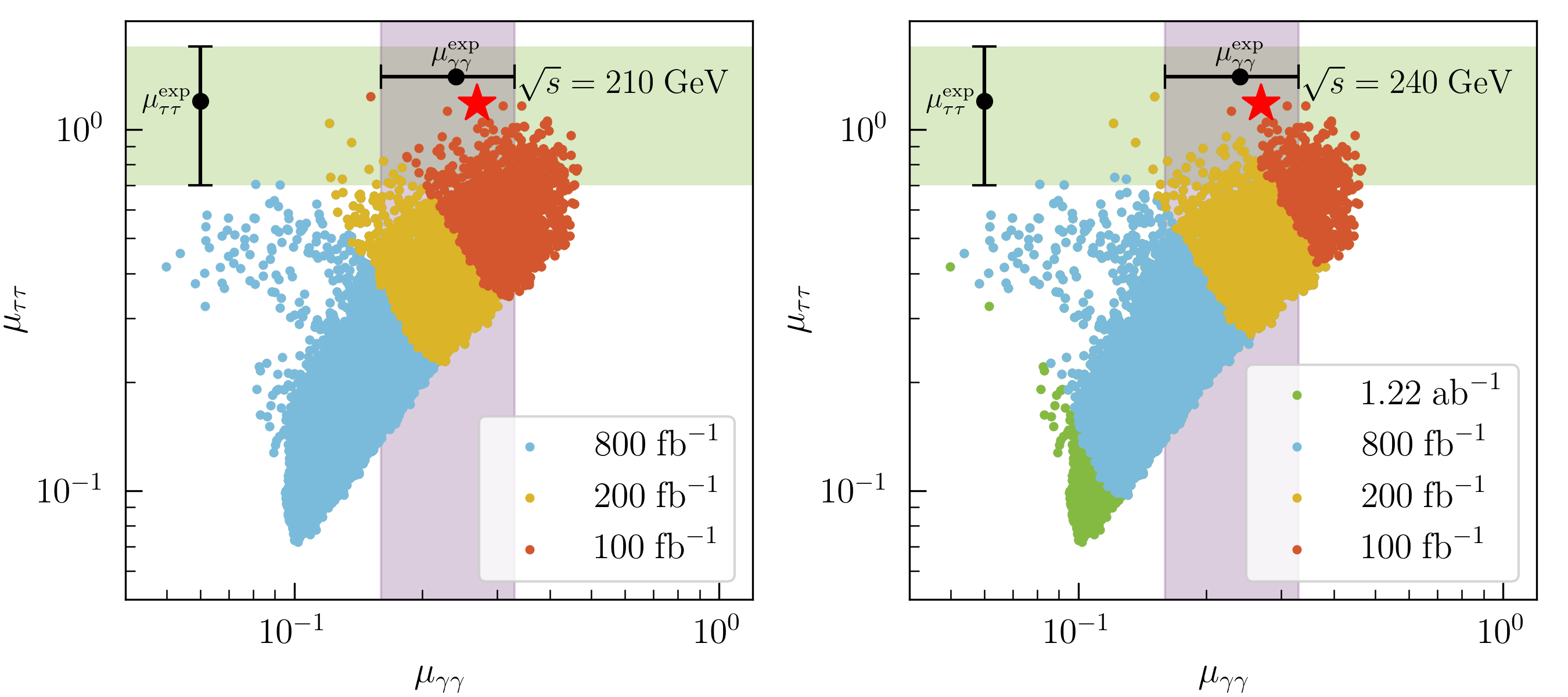}
\caption{\label{f5}
The coverage ability of CEPC for surviving samples in the $\mu_{\tau\tau}$ versus $\mu_{\gamma\gamma}$ plane at $\sqrt{s}=210$ GeV (left panel) and $\sqrt{s}=240$ GeV (right panel).
CEPC with $L=100,~200,~800 \fbm,~\mathrm{and}~1.22\abm$ can cover the red, yellow, blue, and green samples at the $5\sigma$ level, respectively.
All the surviving samples can be covered at $5\sigma$ for $\sqrt{s} = 210$ GeV with $L=800\fbm$ and $\sqrt{s} = 240$ GeV with $L=1.22\abm$.
The red star marks the best-fit point. 
The green and purple shaded bands indicate the experimental $1\sigma$ ranges for the $\tau^+\tau^-$ and $\gamma\gamma$ channels, respectively. 
}
\end{figure*}

After applying the constraints mentioned above, surviving samples with $\chi_{h_{95}}^2<7.82$ are shown in Fig.\ref{f5} in the $\mu_{\tau\tau}$ versus $\mu_{\gamma\gamma}$ plane $\sqrt{s} = 210$ (left) and $240$ (right) GeV.
CEPC with $L=100,~200,~800 \fbm,~\mathrm{and}~1.22\abm$ can cover the red, yellow, blue, and green samples at the $5\sigma$ level, respectively.
All the surviving samples can be covered at $5\sigma$ for $\sqrt{s} = 210$ GeV with $L=800\fbm$ and $\sqrt{s} = 240$ GeV with $L=1.22\abm$.
The green and purple regions in Fig.~\ref{f5} indicate the $1\sigma$ excesses in the $\tau^+\tau^-$ and $\gamma\gamma$ channels, respectively, as reported by experiments.
The best-fit point, marked by a red star, yields $\chi^2_{h_{95}} = ~0.24$, corresponding to a $p$-value of 0.971. For this point, a $2\sigma$ sensitivity can be achieved with $L = 10\,\fbm$ and $5\sigma$ with $L = 58\,\fbm$ at $\sqrt{s} = 210$\,GeV. 
At $\sqrt{s} = 240$\,GeV, the same levels of sensitivity require $L = 13\,\fbm$ and $L =80\,\fbm$, respectively. 
Apart from the sensitivity projections in Fig.~\ref{f5} and associated discussion, our analysis is independent of the specific choice of $\mu_{\tau\tau}$. A smaller consistent $\mu_{\tau\tau}$ would require a correspondingly higher integrated luminosity to maintain equivalent coverage of the parameter space compatible with the current excesses.
These results demonstrate that the $e^+e^- \to Z(\to \mu^+\mu^-) S(\to \tau^+\tau^-)$ channel at CEPC offers excellent potential to probe the 95\,GeV excess in the N2HDM-Flipped framework.

\paragraph*{5. Conclusions and outlook.}
Our study identifies $\sqrt{s}=210$\,GeV as the optimal center-of-mass energy to study the hypothetical scalar with 95 GeV or a slightly higher masses.  
Without ML techniques, the benchmark point requires integrated luminosities of $480\,(690)$\,fb$^{-1}$ to reach $5\sigma$ significance at $\sqrt{s}=210\,(240)$\,GeV. 
A DNN classifier can reduce these numbers to $210\,(310)$\,fb$^{-1}$ at the corresponding energies.  
At $L = 20~\mathrm{ab}^{-1}$, the CEPC's $5\sigma$ sensitivity to the signal strength $\mu_{\tau\tau}^{ZS}$ reaches 0.016 and 0.020 for $\sqrt{s} = 210$ GeV and 240 GeV, respectively. The corresponding thresholds for a 5\% precision measurement are $\mu_{\tau\tau}^{ZS} > 0.10$ and $>0.12$. 
At $\sqrt{s}=210$ GeV (240 GeV), $5\sigma$ coverage of all N2HDM-Flipped samples with $\chi^2_{h_{95}}<7.82$ requires $L=800\ \mathrm{fb}^{-1}$ (1.22 $\mathrm{ab}^{-1}$).
These results suggest that an early CEPC run at 210\,GeV, combined with modern machine-learning selection, offers the most efficient strategy for probing the 95\,GeV excess.  
The analysis framework can be readily adapted to alternative lepton colliders such as the ILC and FCC-ee, and adapted to other light-scalar scenarios.

\paragraph*{Acknowledgement.}
This work was supported by the National Natural Science Foundation of China under Grant No. 12275066 and by the startup research funds of Henan University. 
The work of K. Wang was also supported by the Open Project of the Shanghai Key Laboratory of Particle Physics and Cosmology under Grant No. 22DZ2229013-3. 
And the work of M. Ruan was also supported by the National Key Program for S$\&$T Research and Development under Contract No. 2024YFA1610603. 





\bibliographystyle{apsrev4-1}
\bibliography{apssamp}

\end{document}